\documentclass[preprint]{JHEP3} 

\JHEPspecialurl{http://jhep.sissa.it/JOURNAL/JHEP3.tar.gz}

\usepackage{epsfig,multicol}

\def \sech{\mathop{\rm sech}\nolimits}
\newcommand\fverb{\setbox\fverbbox=\hbox\bgroup\verb}
\newcommand\fverbdo{\egroup\medskip\noindent%
			\fbox{\unhbox\fverbbox}\ }
\newcommand\fverbit{\egroup\item[\fbox{\unhbox\fverbbox}]}
\newbox\fverbbox

\title{Domain walls in a non-linear $\mathbb{S}^2$-sigma model with homogeneous quartic polynomial potential}

\author{A. Alonso-Izquierdo$^{a,c}$, A.J. Balseyro Sebasti\'an, M.A. Gonz\'alez Le\'on$^{a,c}$ \\
$^{a}$ Departamento de Matematica
Aplicada, Universidad de Salamanca, SPAIN
\\$^{c}$ IUFFyM, Universidad de Salamanca, SPAIN
}

\abstract{In this paper the domain wall solutions of a Ginzburg-Landau non-linear $\mathbb{S}^2$-sigma hybrid
model are exactly calculated. There exist two types of basic domain walls and two families of composite domain walls. The domain wall solutions have been identified by using a Bogomolny arrangement in a system of sphero-conical coordinates on the sphere $\mathbb{S}^2$. The stability of all the domain walls is also investigated.}


\begin{document}


\section{Introduction}

Research on topological defects in Field Theory has been an active topic since now more than four decades ago \cite{Rajaraman1982,Rebbi1984,Kolb1990,Vilenkin1994,Manton2004}. These types of solutions when arising in (1+1)-dimensional scalar field theory models are generically called kinks whereas vortices and monopoles are respectively used to name topological defects in the (1+2) and (1+3) dimensional Higgs models. All of them are finite energy solutions of the field equations whose energy densities are localized. In contrast, domain walls are topological defects in (1+d)-dimensional scalar field theory models which carry finite surface tension. In Cosmological models of the early Universe domain walls are formed through the Kibble mechanism when a discrete symmetry is broken at a phase transition; see \cite{Vilenkin1994,Kibble1976}. These solutions are interesting because the topological constraints guarantee its stability, which prevent them from decaying into the vacuum sector. These topological defects also play an important role in  Randall-Sundrum scenarios \cite{Randall1999}, in which space-time is five-dimensional and domain walls correspond to extended 3-branes where the particle dynamics is confined; see \cite{Boonstra1999, Goldberger1999, Csaki2000, DeWolfe2000,Gregory2000, Gremm2000, Gherghetta2000, Campos2002, Cardoso2002, Roessl2002, Melfo2003, Bazeia2004, Barbosa2005, Souza2008, Bazeia2016}. In this context explicit expressions of BPS multi-walls have been identified in some solvable models in ${\cal N}=1$ supergravity \cite{Eto2003, Eto2004}. The formation of domain wall networks and its evolution is also a problem of particular interest in the physical literature. Both numerical approaches \cite{Oliveira2005} and analytic studies based on the moduli space \cite{Eto2007} have been implemented in order to study the low-energy dynamics of these networks. The existence of domain walls in two-component scalar field theory models when the potential involves several minima have been extensively studied. The explicit expressions of BPS kinks or domain walls have been identified in the MSTB model and its generalizations \cite{Montonen1976,Sarker1976,Ito1985,Alonso1998,Alonso2000,Alonso2008}, the BNRT model \cite{Bazeia1995,Bazeia1997,Shifman1998, Alonso2002c}, the Wess-Zumino model \cite{Gibbons1999, Cecotti1993, Saffin1999, Alonso2000b}, etc. Junctions of domain walls, which appear in this type of models are also addressed, see \cite{Oda1999, Carroll2000, Bazeia2000, Binosi2000, Shifman2000}.

In this paper we shall investigate the very rich moduli space of domain walls in a hybrid of the non-linear $\mathbb{S}^2$ sigma model and the Ginzburg-Landau theory of phase transitions. The key feature of these systems is that the target manifold is the sphere ${\mathbb S}^2$. This fact allows these models to have important applications in solid state physics, especially in the so called spintronics. Phenomena such as exchange, anisotropy, and dipole-dipole interactions arising in magnetic materials can be responsible for the existence of several non-collinear ground state configurations in these substances. This situation can give rise to the presence of topological defects describing spin chains in magnetic materials. For example, Haldane constructed a $O(3)$ non-linear sigma field theory model to describe the low-energy dynamics of large-spin one-dimensional Heisenberg antiferromagnets, see \cite{Haldane1983}. In this paper Haldane semiclassically quantizes the soliton type solution of the model. In \cite{Alonso2008b, Alonso2010} the authors obtain the exact expressions of the spin solitary waves in similar problems. The computation of the one-loop mass shifts to the classical masses of these solutions is accomplished in \cite{Alonso2009b}. Other relevant work in this framework is the identification of chiral magnetic soliton lattices present on a chiral helimagnet ${\rm Cr}_{\frac{1}{3}}{\rm NbS}_2$ \cite{Togawa2012}. An investigation of these topologically protected magnetic solitons with applications to logical operations and/or information storage is addressed in \cite{Koumpouras2016}. On the other hand, topological defects in massive non-linear sigma models have also been profusely studied in different supersymmetric models, see \cite{Eto2006, Arai2002, Dorey1998,Naganuma2001b}. A remarkable result in this context is that composite solitons in ${\it d}=3+1$ of Q-strings and domain walls are exact BPS solutions that preserve $\frac{1}{4}$ of the supersymmetries \cite{Gauntlett2001, Isozumi2005}. Skyrmions can also be found in certain non-linear sigma models. Indeed, the original Skyrme model \cite{Skyrme1961} can be understood as a (3+1) dimensional $O(4)$ non-linear sigma model, to which the Skyrme term is added. Skyrmions also exist in 2+1 dimensional models, which are referred to as the $O(3)$ baby-Skyrmion models \cite{Piette1995, Piette1995b, Kudryavtsev1998,Nitta2013}.

In particular, the Ginzburg-Landau massive non-linear ${\mathbb S}^2$-sigma model discussed in this paper includes a non-negative homogeneous quartic polynomial contribution in the fields (modulo the $\mathbb{S}^2$-constraint) to the potential energy density. This term spontaneously breaks the $O(3)$ symmetry of the sigma model down to the $\mathbb{Z}_2 \times \mathbb{Z}_2 \times \mathbb{Z}_2$ symmetry generated by the reflections $\phi_i\rightarrow -\phi_i$ with $i=1,2,3$. As a consequence, six vacuum configurations emerge in this model, which are maximally separated on the sphere $\mathbb{S}^2$. In other words, the ground state configurations point in the cardinal directions (the six principal directions) of the three-dimensional space. This system can be understood as a low-energy limit of an effective field theory derived from a modified Heisenberg model for magnetic crystals. If the crystal structure is simple cubic there are six nearest neighbors (atoms) for any given lattice point. A weak anisotropy, which breaks the $O(3)$ symmetry of the ground state arising in the classical Heisenberg model, can emerge due to interactions of electrons with the neighbor atoms, fixing the vacua along the principal cube directions. In this framework, it seems interesting to investigate possible effective field theories with potentials involving six ground state configurations.

Anisotropy terms of the form $h_{\rm an}=\frac{1}{2} B M_z^2$ have been added to the Heisenberg Hamiltonian to describe the behavior of chiral magnetic soliton lattices, see reference \cite{Koumpouras2016}. Here, $M=(M_x,M_y,M_z)$ is the magnetisation vector field and $B$ is a parameter which measures the anisotropy degree. In \cite{Koumpouras2016} a sine-Gordon equation is derived by the authors in the continuum limit by using additional simplifying assumptions. The analytical solutions of this equation are used to analyse spin configurations in chiral helimagnets. The presence of these solitonic configurations have been experimentally observed on these substances by using Lorentz microscopy and small-angle electron diffraction, see \cite{Togawa2012}. In this paper, the expression $h_{\rm an}=\frac{1}{2} \sum_{i=1}^3 B_{ij} M_i^2 M_j^2$ with $i,j=x,y,z$ is assumed to describe the anisotropy contributions in simple cubic magnetic crystals. This function breaks the original $O(3)$ symmetry of the Heisenberg model but preserves a $\mathbb{Z}_2 \times \mathbb{Z}_2 \times \mathbb{Z}_2$ symmetry. A non-linear $\mathbb{S}^2$-sigma model with potential $V(\vec{\phi})= \frac{1}{2} \sum_{i=1}^3 B_{ij} \phi_i^2 \phi_j^2$ is obtained in the continuum limit. We are interested in investigating this model for the particular choice $B_{23}=B_{13}-B_{12}$, where analytical expressions for the solutions can be extracted. It will be proved that the search for domain walls in this model is tantamount to the search for finite action trajectories in a Neumann type mechanical system \cite{Neumann1859}, which is completely integrable \cite{Moser1980, Dubrovin1981}. Indeed, separation of variables is obtained for the static field equations by using sphero-conical coordinates. This fact allows us to identify the analytical expressions of the whole static domain wall variety. This set comprises two types of basic topological domain walls. The rest of solutions in this set are non-linear combinations of these two single topological defects at static equilibrium. In particular, two different families of composite domain walls are found, which consist of two and four single lumps. Several sum rules are identified connecting the total energy of the different solutions.

The organization of the paper is as follows. In Section 2 we introduce the model and the set of vacua is described. In Section 3 the whole domain wall variety is analytically identified. The exact expressions of the singular domain walls will be established by using spherical coordinates. The two previously mentioned families of composite domain walls are also exactly calculated. On this occasion, the use of sphero-conical coordinates and the Bogomolny approach \cite{Bogomolny1976} allows the identification of first order differential equations describing these solutions, which can be solved. In Section 4 a linear stability study for the found static domain walls is provided. Last Section will be employed to introduce further comments and some suggestions for future lines of enquiry.

\section{The model}

We shall deal with a one-parameter family of $(1+3)$-dimensional $O(3)$ non-linear sigma models, whose dynamics is governed by the action
\begin{equation}
S[\vec{\phi}]= \int d^4 x \, \Big\{ \frac{1}{2}
\partial_\mu\vec{\phi}\cdot \partial^\mu \vec{\phi}
 -V(\vec{\phi};\sigma)\Big\} \hspace{0.2cm}, \label{action}
\end{equation}
where $\vec{\phi}(x^\mu)$ is a map from the Minkowski space $\mathbb{R}^{1,3}$ into the $\mathbb{S}^2$-sphere of radius $R$, $\vec{\phi} : \mathbb{R}^{1,3} \rightarrow \mathbb{S}^2$.
The Minkowski metric tensor is chosen as  $g^{\mu\nu}={\rm diag}(1,-1,-1,-1)$. The fields, space coordinates and parameters introduced in (\ref{action}) are assumed to be dimensionless. Setting an ortho-normal frame $\{\vec{e}_a\}_{a=1,2,3}$ in $\mathbb{R}^3$ with the usual inner product and norm the field $\vec{\phi}(x^\mu)$ is given by
\[
\vec{\phi}(x^\mu) = \sum_{a=1}^3 \phi_a(x^\mu) \vec{e}_a\hspace{0.2cm},
\]
where the field components $\phi_a(x^\mu)$ must comply with the constraint
\begin{equation}
\phi_1^2(x^\mu)+\phi_2^2(x^\mu)+\phi_3^2(x^\mu)=R^2\hspace{0.2cm}. \label{constraint}
\end{equation}

The introduction of a Ginzburg-Landau type potential energy density $V(\vec{\phi};\sigma)$ in the action functional (\ref{action}) breaks the $O(3)$ symmetry of the system and leads to a spontaneous symmetry breaking scenario where domain walls can emerge. An interesting choice of the function $V(\vec{\phi},\sigma)$ is given by the non-negative homogeneous quartic polynomial
\begin{equation}
V(\phi_1,\phi_2,\phi_3;\sigma)= \frac{1}{2} \Big[ \phi_1^2 \,\phi_3^2 + \sigma^4  \,\phi_2^2  \,\phi_3^2 + \bar{\sigma}^4  \,\phi_1^2 \, \phi_2^2 \Big] \hspace{0.2cm}, \label{CartesianPotential}
\end{equation}
where $\bar{\sigma}^2=1-\sigma^2$ and $\sigma\in (0,1)$. The action functional (\ref{action}) is now invariant under the symmetry group $\mathbb{G}=\mathbb{Z}_2 \times \mathbb{Z}_2 \times \mathbb{Z}_2$ generated by the transformations $\pi_i:\phi_i \rightarrow -\phi_i$ with $i=1,2,3$. Moreover, a duality in the family of models can be constructed by simultaneously swapping the parameter values $\sigma \leftrightarrow \bar{\sigma}$ and the field components $\phi_1\leftrightarrow \phi_3$. The study can thus be restricted to the parameter interval $\sigma^2 \in(0,\frac{1}{2}]$ and the results for the rest of cases can be derived by using this duality.

The non-linear field equations of the system, obtained from the action (\ref{action}), are given by
\begin{eqnarray}
\partial_0^2 \phi_1 - \nabla^2 \phi_1 &=&  \phi_1 (\lambda -\bar{\sigma}^4 \phi_2^2- \phi_3^2) \hspace{0.5cm}, \nonumber  \\
\partial_0^2 \phi_2 - \nabla^2 \phi_2 &=& \phi_2 (\lambda -\bar{\sigma}^4 \phi_1^2- \sigma^4 \phi_3^2)\hspace{0.1cm}, \label{pde}  \\
\partial_0^2 \phi_3 - \nabla^2 \phi_3 &=&  \phi_3 (\lambda -\phi_1^2- \sigma^4 \phi_2^2) \hspace{0.5cm}, \nonumber
\end{eqnarray}
where as usual $\nabla^2=\nabla \cdot \nabla$ and $\nabla=\frac{\partial}{\partial x^1} \vec{e}_1 + \frac{\partial}{\partial x^2} \vec{e}_2 + \frac{\partial}{\partial x^3} \vec{e}_3$.
The parameter $\lambda$ is the Lagrange multiplier associated with the constraint (\ref{constraint}), which follows the form
\[
\lambda = \frac{1}{R^2} \sum_{a=1}^3 \left[ -(\partial_0 \phi_a)^2 + \nabla\phi_a \cdot \nabla \phi_a \right] + \frac{4}{R^2} V(\phi_1,\phi_2,\phi_3,\sigma) \hspace{0.2cm} .
\]
The spatial integral of the energy density
\begin{equation}
{\cal E}[\vec{\phi}]= \frac{1}{2} \sum_{\mu=0}^4 \partial_\mu \vec{\phi}\cdot \partial_\mu \vec{\phi} + V(\vec{\phi};\sigma)  \label{energydensity}
\end{equation}
provides the total energy $E$ of the configuration $\vec{\phi}(x^\mu)$, i.e.,
\begin{equation}
E[\vec{\phi}(x^\mu)]= \int_{\mathbb{R}^3} d^3 x \,   {\cal E}[\vec{\phi}(x^\mu) ]  \hspace{0.2cm}. \label{energy}
\end{equation}
It is clear from (\ref{CartesianPotential}) that the set ${\cal M}$ of vacua (zero energy static homogeneous solutions) for our model comprises the six absolute minima of the potential $V(\vec{\phi};\sigma)$
\begin{equation}
{\cal M} = \Big\{  V_{1a}= \Big((-1)^a R,0,0\Big), \, V_{2b}=\Big(0,(-1)^b R,0\Big), \, V_{3c}=\Big(0,0,(-1)^c R\Big)\Big\}\hspace{0.2cm},
\end{equation}
where $a,b,c=0,1$. These points are located at the intersection between the Cartesian axes and the sphere $\mathbb{S}^2$. The six vacua of the system are maximally separated on the sphere. The plane wave expansion around the vacua $V_{i\pm}$ lets us identify the particle spectra in the corresponding quantum theory, which are determined by the mass matrices
\[
M^2(V_{1a}) =R^2 \left( \begin{array}{cc} \bar{\sigma}^4 & 0 \\ 0 & 1  \end{array} \right) \hspace{0.2cm},\hspace{0.2cm} M^2(V_{2b}) = R^2 \left( \begin{array}{cc} \bar{\sigma}^4 & 0 \\ 0 & \sigma^4  \end{array} \right) \hspace{0.2cm},\hspace{0.2cm} M^2(V_{3c}) =R^2 \left( \begin{array}{cc} 1 & 0 \\ 0 & \sigma^4  \end{array} \right) \hspace{0.3cm} .
\]

\section{Variety of domain walls }

Our main goal in this paper is to identify the explicit expressions of the domain wall solutions which arise in this model. \textit{Domain walls} are non-singular solutions of the field equations (\ref{pde}) such that their energy densities have a space-time dependence of the form: ${\cal E}(x^0,x^1,x^2,x^3)={\cal E}(x^1-vx^0)$, where $v$ is a velocity vector in the $x^1$ direction.
In addition, the tension of the wall (2-brane), defined as
\begin{eqnarray}
\Omega(\vec{\phi})&=&\lim_{L\to\infty}\frac{1}{L^2} \int_{-L/2}^{L/2} dx^3 \int_{-L/2}^{L/2}dx^2 \int_{-\infty}^\infty dx^1 {\cal E} [\vec{\phi}]=\nonumber \\ &=& \int dx^1 \Big[ \frac{1}{2}\,  \frac{d \vec{\phi}}{dx^1}\cdot \frac{d\vec{\phi}}{dx^1}\, +\, V(\vec{\phi})\Big] = \int dx^1 \omega(x^1) \hspace{0.2cm}, \label{tension}
\end{eqnarray}
where $L^2$ is the area of a square in the $x_2-x_3$ plane, must be finite. The domain wall tension density is localized around some certain values of the $x^1$ coordinate. Therefore, these solutions can be interpreted as solitonic (thick) 2-branes orthogonal to the $x^1$-axis. Obviously, the choice of the coordinate $x^1$ in the previous definition is a convention and rotations in the spatial axes can be used to set distinct orientations of the domain walls. Solutions which consist of a solitary tension density lump will be referred to as \textit{single domain walls}. On the other hand, the term \textit{composite domain walls} will be used when referring to solutions which can be interpreted as a combination of several single domain walls.

For the sake of simplicity the notation $x^0\equiv t$ and $x^1\equiv x$ is used from now on. The Lorentz invariance of the model implies that it suffices to know the $t$-independent
solutions $\vec{\phi}(x)$ in order to obtain the domain walls of the model:
$\vec{\phi}(t,x)=\vec{\phi}(x-vt)$. Bearing this in mind, the search of domain walls for our model is tantamount to identifying the stationary points of the tension functional (\ref{tension}) belonging to the configuration space ${\cal C}= \{\vec{\phi}: {\rm Maps}(\mathbb{R} , \mathbb{S}^2) /{\rm Maps}({\mathbb R},{\rm point}) : \Omega(\vec{\phi}) < +\infty\}$. These requirements lead to the need of solving the following system of three ordinary differential equations:
\begin{eqnarray}
\frac{d^2 \phi_1}{dx^2}\, &=&  -\phi_1 (\lambda -\bar{\sigma}^4 \phi_2^2- \phi_3^2) \hspace{0.5cm} , \nonumber\\
\frac{d^2 \phi_2}{dx^2}\, &=&  -\phi_2 (\lambda -\bar{\sigma}^4 \phi_1^2- \sigma^4 \phi_3^2) \hspace{0.1cm}, \label{ode} \\
\frac{d^2 \phi_3}{dx^2}\, &=& -\phi_3 (\lambda -\phi_1^2- \sigma^4 \phi_2^2)\hspace{0.5cm} ,
 \nonumber
\end{eqnarray}
under the constraint (\ref{constraint}). The finite wall tension condition is fulfilled if and only if the asymptotic conditions

\begin{equation}
\lim_{x \to \pm \infty} \, \frac{d \vec{\phi}}{dx}\, =\, 0\quad
\mbox{and} \qquad \lim_{x\to \pm \infty} \, \vec{\phi}\, \in {\cal M} \label{asy}
\end{equation}
hold. Thus, the configuration space ${\cal C}$ is the union of 36 disconnected sectors determined by the elements of ${\cal M}$ reached by each configuration at $x \to -\infty$ and $x \to \infty$. Finite tension walls connecting different vacua will be termed as topological walls, whereas non-topological walls refer to solutions belonging to sectors joining the same vacuum.

\subsection{Singular domain walls}

The existence of domain walls whose orbits are pieces of principal great circles (defined by the intersection between the principal planes and the sphere $\mathbb{S}^2$) is investigated in this section. These solutions are called \textit{singular domain walls} because modulo the translational symmetry they are either the only solutions in its topological sector or the extremal member of a continuous family of solutions.

The system of spherical coordinates
\begin{equation}
\phi_1 = \rho \sin \theta \cos \varphi \hspace{0.2cm},\hspace{0.2cm}
\phi_2 = \rho \sin \theta \sin \varphi \hspace{0.2cm},\hspace{0.2cm}
\phi_3 = \rho \cos \theta \hspace{0.2cm},\hspace{0.2cm} \theta\in [0,\pi] , \hspace{0.2cm} \varphi\in (-\pi,\pi] \hspace{0.2cm}, \label{sphericalcoordiates}
\end{equation}
is used to address this problem. The constraint (\ref{constraint}) is immediately satisfied by imposing the restriction $\rho=R$. The evolution equations (\ref{pde}) read
\begin{eqnarray}
&& \frac{\partial^2 \theta}{\partial t^2}-\frac{\partial^2 \theta}{\partial x^2}+\frac{1}{2} \sin(2\theta) \left( \left(\frac{\partial \varphi}{\partial x}\right)^2 - \left(\frac{\partial \varphi}{\partial t}\right)^2 \right) = \nonumber\\
&&= -\frac{R^2}{2} \sin(2\theta) \, \left( \sigma^4+(1-\sigma^4) \cos^2 \varphi -2\sin^2\theta\, \left( \sigma^2+\bar{\sigma}^2 \cos^2\varphi\right)^2\right)\nonumber \\
&& \frac{\partial^2 \varphi}{\partial t^2}-\frac{\partial^2 \varphi}{\partial x^2} +2 \cot \theta \left( \frac{\partial \theta}{\partial t}\frac{\partial \varphi}{\partial t}-\frac{\partial \theta}{\partial x}\frac{\partial \varphi}{\partial x}\right) =\nonumber\\ && =\frac{R^2}{2} \sin (2\varphi) \left( 1-\sigma^4-2\bar{\sigma}^2 \sin^2\theta\, (\sigma^2+\bar{\sigma}^2 \cos^2\varphi)\right)\hspace{0.2cm}, \label{polarequations}
\end{eqnarray}
in the new variables. Rajaraman's trial orbit method \cite{Rajaraman1982} can be successfully applied to the equations (\ref{polarequations}) for the previously mentioned trajectories. We find three types of singular domain walls:


\begin{enumerate}
\item \textit{Equatorial domain walls}: These domain walls connect the four vacua $V_{1a}$ and $V_{2b}$ ($a,b=0,1$) following the equator of the sphere $\mathbb{S}^2$. If the condition $\theta=\frac{\pi}{2}$ is plugged into (\ref{polarequations}) the resulting sine-Gordon equation
    \begin{equation}
    \frac{\partial^2\varphi}{\partial t^2}-\frac{\partial^2\varphi}{\partial x^2}=- \, \bar{\sigma}^4 \frac{R^2}{4} \sin{(4 \varphi)} \label{sG3}
    \end{equation}
    provides us with the profile of this type of solutions. The one-soliton solution of (\ref{sG3}) characterizes the following eight domain walls
     \[
    \theta (x) = \frac{\pi}{2} \hspace{0.3cm},\hspace{0.3cm} \varphi (x) = (-1)^b \cdot \left[a \pi+(-1)^a \arctan \left( e^{q R \bar{\sigma}^2 \, \frac{\bar{x}-vt}{\sqrt{1-v^2}}}\right)\right]  \hspace{0.3cm}, \hspace{0.3cm} \begin{array}{l} q=\pm 1 \\ a,b=0,1 \end{array} \hspace{0.2cm}.
    \]
Lorentz invariance allows us to consider only the case $v=0$ without loss of generality. In the original Cartesian coordinate system the solutions are written as
    \begin{equation}
    \phi_1(x)= \frac{(-1)^a \, R}{\sqrt{1+e^{2  R \bar{\sigma}^2 q \bar{x}}}} \hspace{0.3cm},\hspace{0.3cm} \phi_2(x)=\frac{(-1)^b \, R}{\sqrt{1+e^{- 2 R \bar{\sigma}^2 q \bar{x}}}} \hspace{0.3cm},\hspace{0.3cm} \phi_3(x)=0 \hspace{0.2cm}. \label{EquatorialDW}
    \end{equation}
Here, $\bar{x}$ stands for $\bar{x}=x-x_0$ where $x_0\in \mathbb{R}$ is the domain wall center, and $a, b=0, 1$. The value of $q$ distinguishes between walls and anti-walls. The translational symmetries and spatial parity underlie the presence of the parameters $x_0$ and $q$ in (\ref{EquatorialDW}). The quadrant of the equator where the topological defect is defined is labeled by the pair $(a,b)$. Indeed, a particular solution in (\ref{EquatorialDW}) asymptotically joins the vacuum points $V_{1a}$ and $V_{2b}$. For the sake of conciseness the equatorial domain walls will be denoted as $K_1^{(q,a,b)}(\overline{x})$.
The tension density of the $K_1^{(q,a,b)}(\overline{x})$ solutions
    \[
    \omega[K_1^{(q,a,b)}(\overline{x})]= \frac{R^4 \bar{\sigma}^4}{4} \sech^2 \left( R\,q\, \overline{\sigma}^2 \, \overline{x}\,\right)
    \]
    is confined to a small region, see Figure 1. Therefore, the solutions $K_1^{(q,a,b)}(\overline{x})$ describe single solitary thick 2-branes. This fact is pointed out in the previous notation by means of the subscript 1. The total wall tension amounts to:
    \[
    \Omega[K_1^{(q,a,b)}(\overline{x})]= \frac{1}{2} \, R^3 \,\overline{\sigma}^2 \hspace{0.2cm}.
    \]
In addition to the previously described single domain walls, the multisoliton and breather solutions of the equations (\ref{sG3}) can be exploited to construct complex evolving domain walls. We will not explore this type of solutions in this paper because we are interested in studying the structure of the domain wall variety coming from the two-dimensionality of the internal space.

\FIGURE{
\includegraphics[height=3.cm]{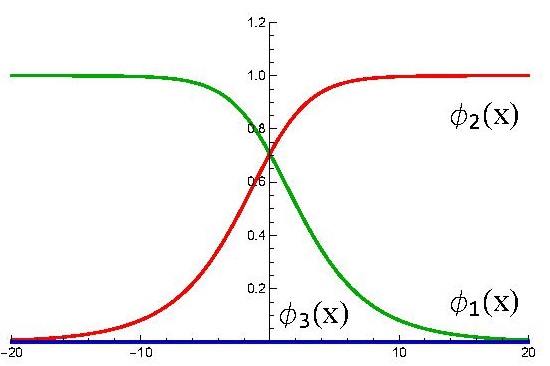}\qquad  \includegraphics[height=3.cm]{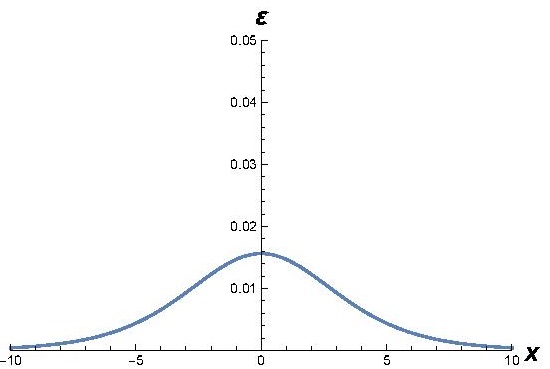}\qquad  
\includegraphics[height=3.cm]{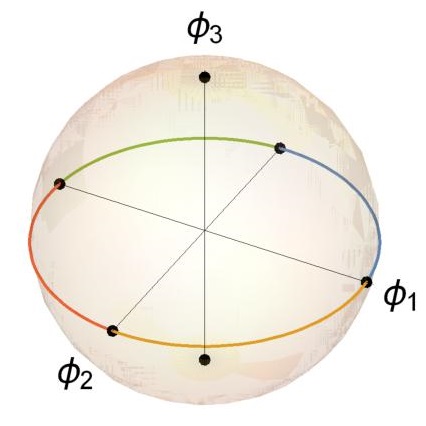}
\caption{Profile of the Cartesian field components (left), wall tension density (middle) and orbits for the $K_1^{(q,a,b)}(\overline{x})$ domain walls (right).}}

\item \textit{($\pm \frac{\pi}{2}$)-Meridian domain walls}: In this case the trial orbit, which will be inserted into the equations (\ref{polarequations}), is the great circle formed by the meridians with longitude $\varphi=\pm \frac{\pi}{2}$. Now, the behavior of the polar variable $\theta$ is set by the sine-Gordon equation
 \begin{equation}
\frac{\partial^2\theta}{\partial t^2}-\frac{\partial^2\theta}{\partial x^2}=  -\frac{R^2\sigma^4}{4} \sin (4 \theta) \hspace{0.2cm}, \label{sG1}
\end{equation}
which leads to the one-soliton domain walls
\[
\theta (x) =  c \pi+(-1)^c \arctan \left(  e^{R \, \sigma^2 \, q \, \bar{x}}\right) \hspace{0.3cm},\hspace{0.3cm} \varphi (x) =  \frac{(-1)^b \pi}{2} \hspace{0.5cm},\hspace{0.5cm} \begin{array}{l} q=\pm 1 \\ b,c=0,1 \end{array} \hspace{0.2cm}.
\]
These eight topological defects, which join the vacuum points $V_{2b}$ and $V_{3c}$ with $b,c=0, 1$, read
\[
\phi_1(x)= 0 \hspace{0.3cm},\hspace{0.3cm} \phi_2(x)= \frac{(-1)^b R}{\sqrt{1+e^{-2 R \sigma^2 \, q \overline{x}}}} \hspace{0.3cm},\hspace{0.3cm} \phi_3(x)=\frac{(-1)^c R}{\sqrt{1+e^{2 R \sigma^2 q \overline{x}}}} \hspace{0.2cm},
\]
in the original variables. These solutions will be denoted as $\overline{K}_1^{(q,b,c)}(\overline{x})$ where the parameters $q$, $b$ and $c$ play the same role as in the previous case. The total wall tension carried by these solutions
\[
\Omega[\overline{K}_1^{(q,b,c)}(\overline{x})]= \frac{1}{2} R^3 \sigma^2
\]
is obtained by integrating the domain wall tension density
\[
\omega[\overline{K}_1^{(q,b,c)}(\overline{x})]=\frac{R^4 \sigma^4 }{4} \sech^2 \left( Rq\sigma^2 \bar{x} \right)
\]
along the $x^1$-coordinate. The distribution of this density implies that the $\overline{K}_1^{(q,b,c)}(\overline{x})$-walls are single thick 2-branes, see Figure 2.

\FIGURE{ \includegraphics[height=3cm]{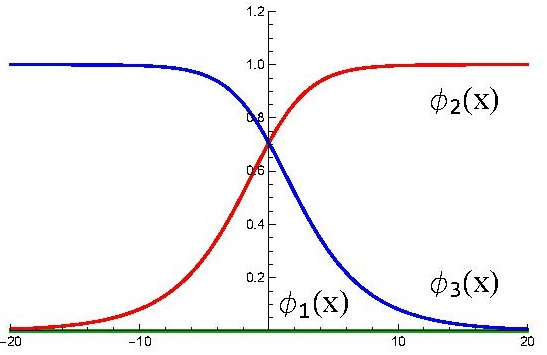} \qquad \includegraphics[height=3cm]{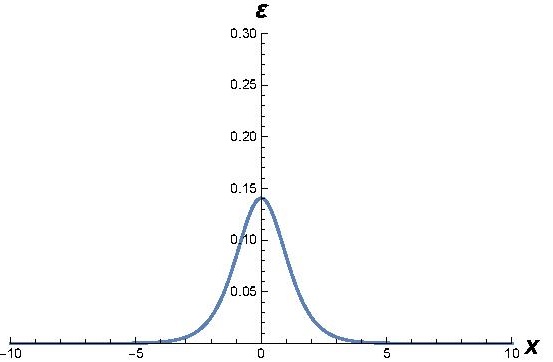} \qquad 
\includegraphics[height=3cm]{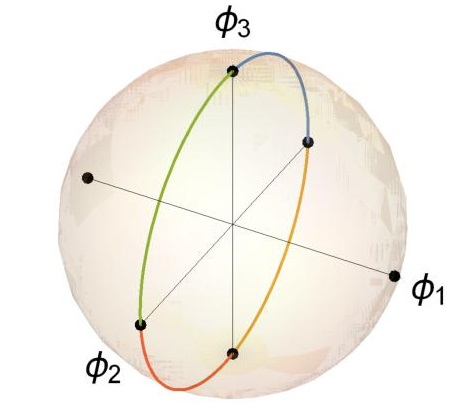} \caption{Profile of the Cartesian field components (left), wall tension density (middle) and orbits for the $\overline{K}_1^{(q,b,c)}(\overline{x})$ domain walls (right).}}

Multisoliton domain walls can also be identified in this context by using different solutions of the sine-Gordon equation.

\item \textit{Prime Meridian domain walls}: The orbits of this type of solutions are defined on the orthodrome composed of the meridian with azimuthal angle $\varphi=0$ and its antimeridian. These trial orbits lead again to a sine-Gordon equation
    \begin{equation}
    \frac{\partial^2\theta}{\partial t^2}-\frac{\partial^2\theta}{\partial x^2}= -\frac{R^2}{4} \sin (4 \theta) \hspace{0.2cm},
    \end{equation}
    which provides us with the eight domain wall solutions
      \[
    \theta (x) =  c \pi+ (-1)^c\arctan \left( e^{R \, q \, \bar{x}}\right) \hspace{0.3cm},\hspace{0.3cm} \varphi (x) = a \pi, \hspace{0.9cm} a,c=0,1 \hspace{0.2cm}.
    \]
In Cartesian coordinates these topological defects follow the form
   \begin{equation}
\phi_1= \frac{(-1)^a R}{\sqrt{1+e^{-2 R q\, \overline{x}}}}, \qquad \phi_2=0,\qquad \phi_3=\frac{(-1)^c R}{\sqrt{1+e^{2 R q \overline{x}}}}  \hspace{0.2cm} . \label{PrimeMeridoanDW}
\end{equation}
   As before, $q=\pm 1$ distinguishes between walls and anti-walls. On the other hand, $a,c=0, 1$ determine if the solution lives in the Northern or Southern Hemisphere and in the Western or Eastern Hemisphere, respectively.

The solutions (\ref{PrimeMeridoanDW}) connect the four vacua $V_{1a}$ and $V_{3c}$ and will be represented by the symbol $K_2^{(q,a,c)}(\overline{x})$. The $K_2^{(q,a,c)}(\overline{x})$ domain wall tension density
\begin{equation}
\omega[K_2^{(q,a,c)}(\overline{x})]=\frac{R^4}{4} \sech^2 \left( R\,q \,\overline{x} \right)
\end{equation}
is concentrated around the wall center $x=x_0$. However, these solutions are the extremal members of a continuous family of composite domain walls and can be interpreted as the overlap of the two solitary wall tension lumps $K_1^{(q,a,b)}(\overline{x})$ and $\overline{K}_1^{(q,b,c)}(\overline{x})$ previously described, see Figure 3. This fact is supported by the energy sum rule
   \begin{equation}
   \Omega [K_2^{(q,a,c)}(x)] =\Omega[K_1^{(q,a,b)}(\overline{x})] +\Omega[\overline{K}_1^{(q,b,c)}(\overline{x})] =  \frac{1}{2} \, R^3 \label{sumrule0}
   \end{equation}
   and will be analytically justified in next sections.

\FIGURE{ \includegraphics[height=3cm]{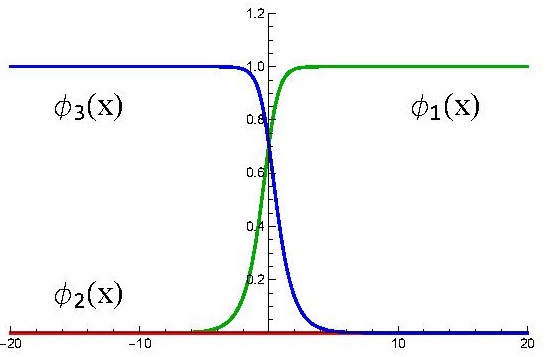}\qquad  \includegraphics[height=3cm]{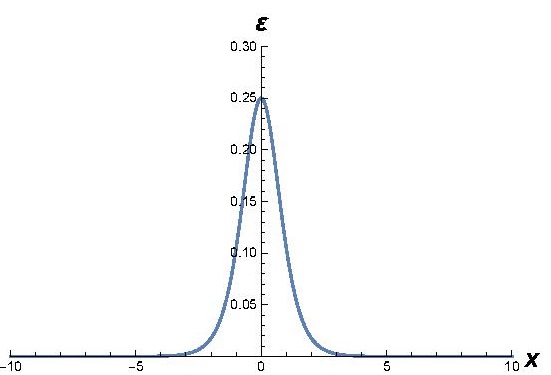}\qquad  
\includegraphics[height=3cm]{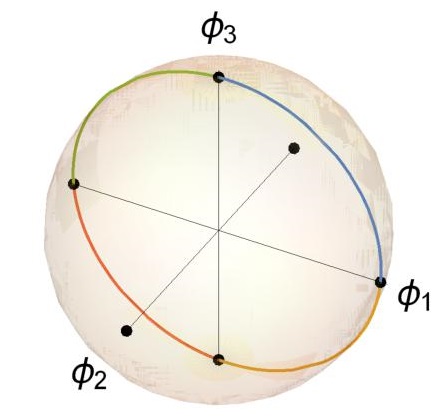} \caption{Profile of the Cartesian field components (left), wall tension density (middle) and orbits for the $K_2^{(q,a,c)}(\overline{x})$ domain walls (right).}}

\end{enumerate}

\subsection{Families of composite domain walls}

In this section the description of the static domain wall manifold for this non-linear $\mathbb{S}^2$-sigma model is completed. We find two one-parameter families of composite domain walls, whose orbits are dense in the sphere $\mathbb{S}^2$. In order to deal with this problem, sphero-conical coordinates on the sphere
\begin{equation}
\phi_1^2=\frac{R^2}{\bar{\sigma}^2} \lambda_1 \lambda_2 \hspace{0.2cm},\hspace{0.2cm} \phi_2^2 = \frac{R^2}{\sigma^2\bar{\sigma}^2} (\bar{\sigma}^2-\lambda_1)(\lambda_2 -\bar{\sigma}^2) \hspace{0.2cm},\hspace{0.2cm} \phi_3^2 = \frac{R^2}{\sigma^2} (1-\lambda_1)(1-\lambda_2)
\label{spheroconical}
\end{equation}
are introduced. The range of the coordinates $\lambda_1$ and $\lambda_2$ in (\ref{spheroconical}) is given by the rectangle $C\equiv \{ (\lambda_1,\lambda_2):0<\lambda_1<\bar{\sigma}^2 <\lambda_2 <1\}$. The map (\ref{spheroconical}) is thus a coordinate chart between $C$ and an octant of the $\mathbb{S}^2$. The change of coordinates is eight-to-one and eight charts are demanded to cover the whole sphere. Piecewise solutions can be constructed by demanding smoothness in the transitions between octants.

The four vertices of the rectangle $C$ are distinguished points in our model. The corner $(\lambda_1,\lambda_2)=(0,\bar{\sigma}^2)$ corresponds to the vacua $V_{3c}$, the point $(\lambda_1,\lambda_2)=(0,1)$ is mapped to the vacua $V_{2b}$ and $(\lambda_1,\lambda_2)=(\bar{\sigma}^2,1)$ goes to the points $V_{1a}$, where $a,b,c=0,1$. The remaining vertex $(\lambda_1,\lambda_2)=(\bar{\sigma}^2,\bar{\sigma}^2)$ represents the four points
\[
F_{ac}= \left( (-1)^a \, R \, \bar{\sigma} , 0, (-1)^c \, R \,\sigma \right) \hspace{0.5cm} a,c=0,1
\]
in Cartesian coordinates. They are the foci of the \lq\lq elliptical" curves defined by the sphero-conical coordinate isolines. In summary, the vacua of our model together with the foci of the coordinate curves of the sphero-conical coordinate system are mapped to the vertices of $C$. The principal great circles become straight lines in the sphero-conical coordinate plane. In particular, the equator is represented by the segment $E=\{(\lambda_1,1): 0< \lambda_1<\bar{\sigma}^2\}$, the $(\pm \frac{\pi}{2})$ meridians are characterized by $E=\{(0,\lambda_2): \bar{\sigma}^2< \lambda_2<1\}$ and the Prime meridian and its antimeridian by the concatenation of the remaining edges $\{(\lambda_1,\bar{\sigma}^2):0< \lambda_1<\bar{\sigma}^2 \} \cup\{ (\bar{\sigma}^2, \lambda_2 ):\bar{\sigma}^2< \lambda_2<1 \}$.

The wall tension can be written as
\[
\Omega(\lambda_1,\lambda_2)= \int dx \Big[ \frac{1}{2} \,g_{11}(\lambda_1,\lambda_2) \Big( \frac{d\lambda_1}{dx} \Big)^2 + \frac{1}{2} \, g_{22}(\lambda_1,\lambda_2) \Big( \frac{d\lambda_2}{dx} \Big)^2 + V (\lambda_1,\lambda_2) \Big]
\]
in the new variables where the metric factors $g=(g^{ij})$ are defined as
\begin{eqnarray*}
g^{11}(\lambda_1,\lambda_2)&=& g_{11}^{-1} (\lambda_1,\lambda_2) =\frac{4\lambda_1(\bar{\sigma}^2- \lambda_1)(1-\lambda_1)}{R^2(\lambda_2-\lambda_1)} \hspace{0.4cm}, \\
g^{22}(\lambda_1,\lambda_2) &=& g_{22}^{-1} (\lambda_1,\lambda_2) = \frac{4\lambda_2(\lambda_2-\bar{\sigma}^2 )(1-\lambda_2)}{R^2(\lambda_2-\lambda_1)} \hspace{0.4cm} .
\end{eqnarray*}
The reason behind the use of the sphero-conical coordinates is unveiled when writing the potential energy density in these variables
\[
V(\lambda_1,\lambda_2)=\frac{R^4}{2 (\lambda_2-\lambda_1)} \Big[\lambda_1 (1-\lambda_1)(\bar{\sigma}^2 - \lambda_1) +\lambda_2 (1-\lambda_2)(\lambda_2 -\bar{\sigma}^2)\Big] \hspace{0.4cm} .
\]
A Bogomolny arrangement can now be applied to the expression of the wall tension $\Omega (\lambda_1,\lambda_2)$, which can be written as
\begin{eqnarray}
\Omega(\lambda_1,\lambda_2)&=& \int dx \Big[ \frac{1}{2} \sum_{i=1}^2 g_{ii} \Big( \frac{d\lambda_i}{dx} \Big)^2 +\frac{1}{2} \sum_{i=1}^2 g^{ii} \Big( \frac{\partial W}{\partial \lambda_i}\Big)^2 \Big] = \nonumber \\
&=& \int dx \, \frac{1}{2} \, \sum_{i=1}^2 g_{ii} \Big( \frac{d\lambda_i}{dx} - g^{ii} \frac{\partial W}{\partial \lambda_i} \Big)^2 + \Big| \int dx \sum_{i=1}^2 \frac{\partial W}{\partial \lambda_i} \frac{\partial \lambda_i}{dx} \Big| \hspace{0.4cm}, \label{bogo}
\end{eqnarray}
where the superpotential $W(\lambda_1,\lambda_2)$ is chosen as
\[
W(\lambda_1,\lambda_2)= \frac{1}{2} R^3 (-1)^\alpha [\lambda_1 + (-1)^\beta \lambda_2] \hspace{0.6cm} \mbox{with} \hspace{0.6cm} \alpha,\beta=0,1.
\]
For configurations belonging to the space ${\cal C}$, the last term in (\ref{bogo})
\[
T= \Big|\int dx \sum_{i=1}^2 \frac{\partial W}{\partial \lambda_i} \frac{\partial \lambda_i}{dx}\Big|
\]
is a topological charge, which is conserved during the evolution of the configuration. Therefore, the static domain walls, which are stationary points of the wall tension functional $\Omega(\lambda_1,\lambda_2)$, must comply with the system of first order differential equations
\begin{eqnarray}
\frac{d\lambda_1}{dx} &=& g^{11} \frac{\partial W}{\partial \lambda_1} = (-1)^{\alpha} \, \frac{2 R\,  \lambda_1 (1-\lambda_1)(\bar{\sigma}^2-\lambda_1)}{\lambda_2-\lambda_1} \hspace{0.8cm}, \label{spcoeq1} \\
\frac{d\lambda_2}{dx} &=& g^{22} \frac{\partial W}{\partial \lambda_2} = \, (-1)^{\alpha+\beta} \frac{2R \, \lambda_2 (1-\lambda_2)(\lambda_2 -\bar{\sigma}^2)}{\lambda_2-\lambda_1} \hspace{0.4cm}, \label{spcoeq2}
\end{eqnarray}
which can be solved for the values $\alpha,\beta=0,1$. The integration of the equations (\ref{spcoeq1}) and (\ref{spcoeq2}) leads to the expressions
\begin{eqnarray}
&& \frac{\bar{\sigma}^2 - \lambda_1}{\lambda_1^{\sigma^2} (1-\lambda_1)^{\bar{\sigma}^2}}\cdot \Big[\frac{\lambda_2 -\bar{\sigma}^2}{\lambda_2^{\sigma^2} (1-\lambda_2)^{\bar{\sigma}^2}}\Big]^{(-1)^{\beta}} = e^{2 R \sigma^2 \gamma} \hspace{0.4cm}, \label{spcosol1} \\
&& \frac{\bar{\sigma}^2 - \lambda_1}{1-\lambda_1} \cdot\Big[\frac{\lambda_2 -\bar{\sigma}^2}{1-\lambda_2}\Big]^{(-1)^{\beta}} =e^{(-1)^{\alpha} 2 R \sigma^2 \overline{x}} \hspace{2.1cm}. \label{spcosol2}
\end{eqnarray}
The pair of relations (\ref{spcosol1}) and (\ref{spcosol2}) determine two distinct families of domain walls $\vec{\phi}(\overline{x},\gamma)$ parameterized by the value of the integration constant $\gamma \in \mathbb{R}$ and $\beta=0,1$. The relation (\ref{spcosol1}) sets the domain wall orbits whereas (\ref{spcosol2}) determines the spatial dependence of the solution. Walls and antiwalls are obtained as solutions of (\ref{spcoeq1}) and (\ref{spcoeq2}) with different values of $\alpha$. A direct manipulation of the equations (\ref{spcosol1}) and (\ref{spcosol2}) allows us to write them in a simpler form:
\begin{eqnarray}
&& \frac{\bar{\sigma}^2 - \lambda_1}{1-\lambda_1} \cdot \Big[\frac{\lambda_2 -\bar{\sigma}^2}{1-\lambda_2}\Big]^{(-1)^{\beta}}=e^{(-1)^{\alpha} 2 R \sigma^2 \overline{x}} = A(x) \hspace{0.4cm}, \label{spcosol3} \\
&& \frac{\lambda_1}{1-\lambda_1} \cdot \Big[\frac{\lambda_2}{1-\lambda_2}\Big]^{(-1)^{\beta}}=e^{ 2 R [(-1)^{\alpha}\overline{x}-\gamma]} =B(x) \hspace{0.3cm}. \label{spcosol4}
\end{eqnarray}
where for sake of simplicity in subsequent expressions we have introduce the functions $A(x)$ and $B(x)$ to denote the exponentials in (\ref{spcosol3}) and (\ref{spcosol4}).

In summary, the relations (\ref{spcosol1}) and (\ref{spcosol2}), or its equivalent ones (\ref{spcosol3}) and (\ref{spcosol4}), determine two one-parameter families of domain walls distinguished by the value of $\beta$. We shall discuss separately each of these families in the following subsections.

\subsubsection{Family of two-lump composite domain walls}

In this section the domain walls extracted from (\ref{spcosol3}) and (\ref{spcosol4}) with $\beta=0$ are described. It is a straightforward, albeit tedious, computation to check that
\begin{eqnarray*}
\lambda_i(x) &=& \frac{1}{2 (A(x)+\bar{\sigma}^2 +B(x) \sigma^2)} \left( A(x)+\bar{\sigma}^4 +(1-\bar{\sigma}^4) B(x)+ \right. \\ && \left.  + (-1)^i \sqrt{A(x)^2+2 A(x) \bar{\sigma}^4 +\bar{\sigma}^8+2 A(x) B(x) \sigma^4 -2 B(x) \bar{\sigma}^4 \sigma^4 + B(x)^2 \sigma^8} \right)
\end{eqnarray*}

\FIGURE{\includegraphics[height=2.5cm]{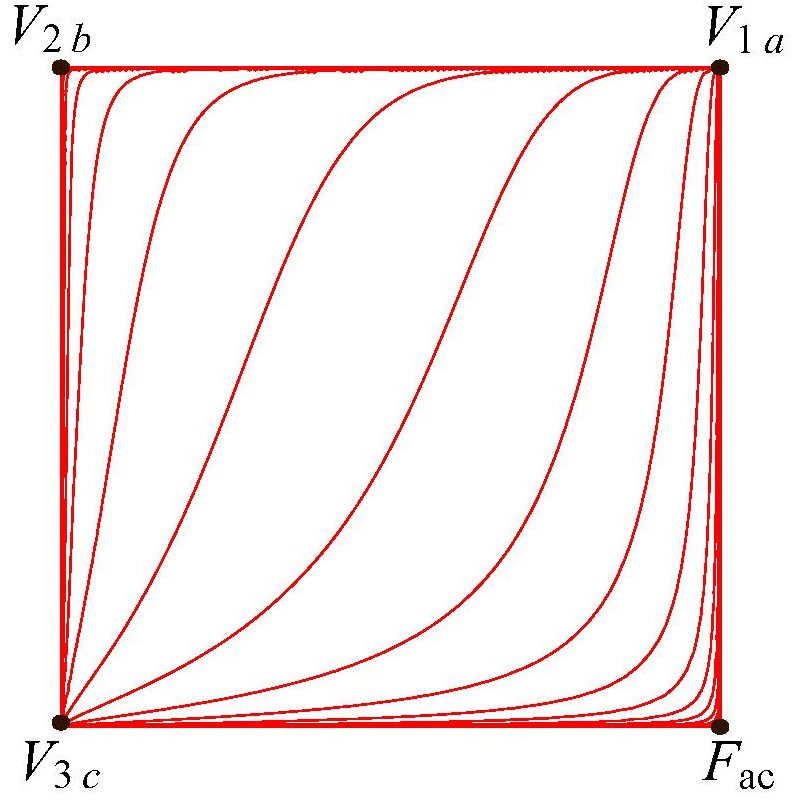} \hspace{0.2cm} 
\includegraphics[height=2.5cm]{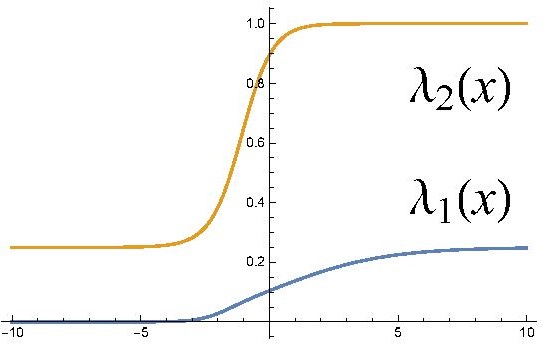} \caption{Orbits for several members of the family $K_2^{(q,a,b,c)}(\bar{x},\gamma)$ (left) and field components (right) graphically represented in sphero-conical coordinates.}}

\noindent where $i=1,2$. In Figure 4 the domain wall orbits in the sphero-conical plane have been depicted for several values of $\gamma$ together with the profile of its components. This type of solutions asymptotically connects the vacua $V_{3c}$ and $V_{1a}$ with $a,c=0,1$ and are confined to a sphere octant. Notice that the components $\lambda_1(x)$ and $\lambda_2(x)$ takes values on the corresponding ranges of the sphero-conical coordinates. In the Cartesian coordinates, the domain walls follow the form
\begin{eqnarray}
\phi_1^{(a,b,c)}(x,\gamma) &=&(-1)^a R \sigma \, \sqrt{\frac{B(x)}{A(x)+\bar{\sigma}^2 +B(x) \sigma^2}} \hspace{0.3cm},  \nonumber \\
\phi_2^{(a,b,c)}(x,\gamma) &=&(-1)^b R \, \sqrt{\frac{A(x)}{A(x)+\bar{\sigma}^2 +B(x) \sigma^2}} \hspace{0.3cm}, \label{family00}\\
\phi_3^{(a,b,c)}(x,\gamma) &=&(-1)^c R \bar{\sigma} \, \frac{1}{\sqrt{A(x)+\bar{\sigma}^2 +B(x) \sigma^2}} \hspace{0.3cm}, \nonumber
\end{eqnarray}
where the values $a,b,c=0,1$ determine the sphere octant where these domain walls are confined. In Figure 5 the orbits of several members of this family have been plotted on the sphere $\mathbb{S}^2$. In this figure the graphics of the Cartesian field components and the domain wall tension densities have also been included.

\FIGURE{\includegraphics[height=3cm]{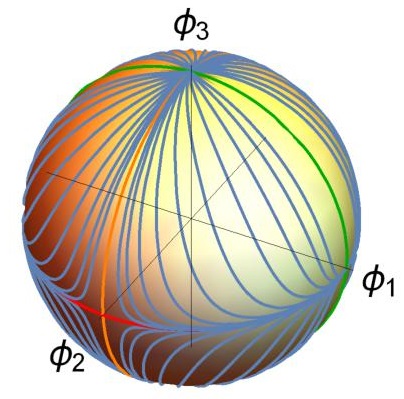} \hspace{0.4cm} 
\includegraphics[height=3cm]{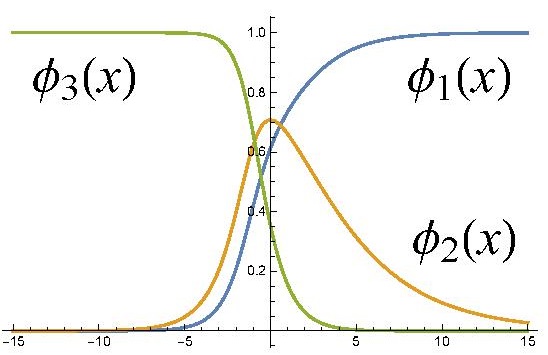} \hspace{0.4cm}  \includegraphics[height=3cm]{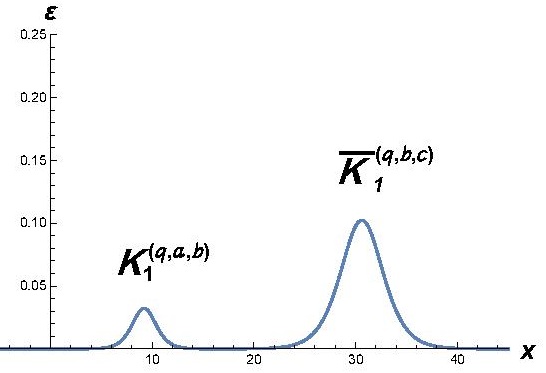}\caption{Orbits for several members of the family $K_2^{(q,a,b,c)}(\bar{x},\gamma)$ (left), field components (middle) and wall tension densities (right) graphically represented in Cartesian coordinates.}}

We shall denote these one-parameter families of domain walls as $K_2^{(q,a,b,c)}(\bar{x},\gamma)$, where as before $q=\pm 1$ distinguishes between walls and antiwalls. The distribution of the tension densities shows that the members of this family can be interpreted as the combination of two separated single domain walls $K_1^{(q,a,b)}(\overline{x})$ and $\overline{K}_1^{(q,b,c)}(\overline{x})$, where the parameter $\gamma$ measures the distance between these tension density lumps. In a more technical way, the single domain walls can be obtained from the family (\ref{family00}) by taking specific limits in the parameter space $(\gamma,x_0)$. It can be checked that
\[
\lim_{\tiny\begin{array}{c} \gamma \rightarrow -\infty \\ \gamma +(-1)^\alpha \overline{\sigma}^2 x_0 \equiv {\rm constant} \end{array}}  K_2^{(q,a,b,c)}(x,\gamma) = K_2^{(q,a,c)}(x) \hspace{0.3cm},
\]
which means that the singular domain walls $K_2^{(q,a,c)}(x)$ introduced in the previous section are extremal members of the one-parameter family $K_2^{(q,a,b,c)}(x,\gamma)$. On the other hand,
\begin{eqnarray*}
\lim_{\tiny\begin{array}{c} \gamma\rightarrow \infty \\ x_0 \,\, {\rm constant} \end{array}}  K_2^{(q,a,b,c)}(x,\gamma)& = &\overline{K}_1^{(q,b,c)}(x) \hspace{0.3cm}, \\
\lim_{\tiny\begin{array}{c} \gamma\rightarrow \infty \\ \gamma+(-1)^\alpha x_0 \equiv {\rm constant} \end{array}}  K_2^{(q,a,b,c)}(x,\gamma)& = &K_1^{(q,a,b)}(x)\hspace{0.3cm}.
\end{eqnarray*}
The energy sum rule (\ref{sumrule0}) is now extended as
\[
\Omega [K_2^{(q,a,b,c)}(\bar{x},\gamma)] =\Omega [K_2^{(q,a,c)}(\bar{x})] =\Omega[K_1^{(q,a,b)}(\overline{x})] +\Omega[\overline{K}_1^{(q,b,c)}(\overline{x})] =  \frac{1}{2} \, R^3 \hspace{0.3cm}.
\]
In summary, the family of $K_2^{(q,a,b,c)}(\bar{x},\gamma)$-domain walls can be understood as two solitonic thick 2-branes which are separated by a distance fixed by the value of the family parameter $\gamma$. The singular member $K_2^{(q,a,c)}(\bar{x})$ arises when the two previously mentioned solitary domain walls are maximally overlapped.

\subsubsection{Family of four-lump composite domain walls}

In order to finish the identification of the static domain wall manifold the equations (\ref{spcosol3}) and (\ref{spcosol4}) with $\beta=1$ must be solved for the sphero-conical variables. The spatial dependence of these variables is given by the expressions
\[
\lambda_1(x)= \frac{B(x) (1+A(x)) \bar{\sigma}^2}{B(x)+A(x) B(x) \bar{\sigma}^2 + A(x) \sigma^2} \hspace{0.3cm}, \hspace{0.3cm} \lambda_2(x)= \frac{ (1+A(x)) \bar{\sigma}^2}{A(x)+ \bar{\sigma}^2 + B(x) \sigma^2}\hspace{0.3cm},
\]

\FIGURE{\includegraphics[height=2.5cm]{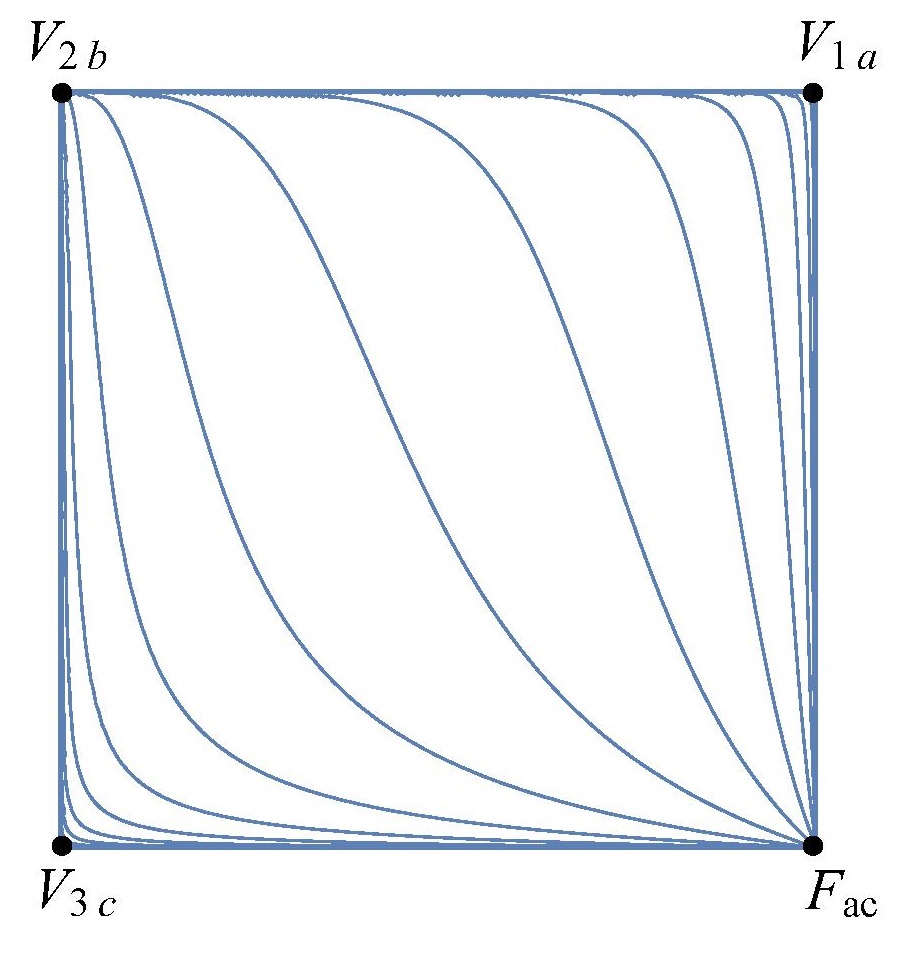} \hspace{0.3cm} 
\includegraphics[height=2.5cm]{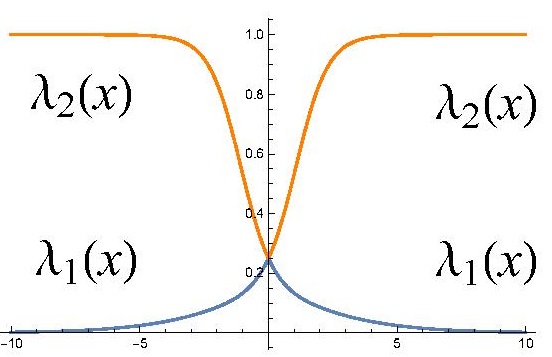} \caption{Orbits for several members of the family $K_4^{(q,a,b,c)}(\bar{x},\gamma)$ (left) and field components (right) graphically represented in sphero-conical coordinates.}}

The trajectories and the field components of this type of solutions are graphically represented in the sphero-conical plane for several values of $\gamma$ in Figure 6. In the sphero-conical rectangle $C$ all the orbits begin at the vertex $V_{2b}$ and monotonically reach the foci $F_{ac}$. Indeed a particular solution arrives at $F_{ac}$ when $\overline{x}=(-1)^\alpha \gamma$, where $\alpha$ distinguishes between walls and antiwalls. In the original fields the tangent vector at the foci $F_{ac}$ is
\[
\frac{d\vec{\phi}}{dx} =(-1)^\alpha R^2 \sigma \bar{\sigma} \Big((-1)^a \sigma \tanh (\sigma^2 R \gamma), - {\rm sign}(\phi_2) \,{\rm sech} (\bar{\sigma}^2 R \gamma) , -(-1)^c \tanh (\sigma^2 R \gamma ) \Big)
\]

At this point in $C$ the solutions carry non-vanishing potential tension density and continue to the next sphere octant represented by the same rectangle $C$. In order to complete the domain wall orbit a new piece of trajectory is needed, which must be chosen in such a way that the smoothness of the solutions is guaranteed. From the previous expression it is clear that this can be achieved by concatenating the previous solution with the one obtained from the equations (\ref{spcosol1}) and (\ref{spcosol2}) with the opposite value of $\alpha$ and $\gamma$. Despite this fact, the global expression

{\small\begin{eqnarray*}
\phi_1 ^{(a,b,c)}(x,\gamma) &=& (-1)^a R \bar{\sigma} (1+ A(x))  \sqrt{\frac{B(x)}{(B(x)+A(x) B(x) \bar{\sigma}^2 + A(x) \sigma^2)(A(x)+\bar{\sigma}^2 + B(x) \sigma^2)}} ,\\
\phi_2^{(a,b,c)}(x,\gamma) &=& (-1)^b  R \sigma \bar{\sigma} (1-B(x)) \sqrt{\frac{A(x)}{(B(x)+A(x) B(x) \bar{\sigma}^2 + A(x) \sigma^2)(A(x)+\bar{\sigma}^2 + B(x) \sigma^2)}}, \\
\phi_3^{(a,b,c)}(x,\gamma)  &=& (-1)^c R \sigma (A(x)+B(x)) \frac{1}{\sqrt{(B(x)+A(x) B(x) \bar{\sigma}^2 + A(x) \sigma^2)(A(x)+\bar{\sigma}^2 + B(x) \sigma^2)}},
\end{eqnarray*}}
can be found for the field components in the Cartesian coordinates. All the orbits asymptotically start at a vacuum point $V_{2b}$ ($b=0,1$), later cross the Prime Meridian, passing to a new sphere octant through one of the foci $F_{ac}$ and asymptotically arrive at the antipodal vacuum, see Figure 7. We shall use the notation $K_4^{(q,a,b,c)}(\overline{x},\gamma)$ to denote this family of topological defects. The wall tension distribution is also illustrated in Figure 7. These domain walls carry four tension lumps.  In particular, the $K_4^{(q,a,b,c)}(\overline{x},\gamma)$ solutions consist of a non-linear combination of a single $\overline{K}_1^{(q,b,c)}(\overline{x})$ and a single $K_1^{(q,a,\overline{b})}(\overline{x})$ domain walls, which are separated by a $K_2^{(q,a,c)}(\overline{x})$-meridian domain wall in the middle, see Figure 7, where $\overline{b}=(b+1){\rm mod}\,2$. In addition, the central lump $K_2^{(q,a,c)}(\overline{x})$ can be understood as the overlapping of the two single walls $\overline{K}_1^{*(q,b,c)}(\overline{x})$ and $K_1^{(q,a,b)}(\overline{x})$ or $\overline{K}_1^{*(q,\overline{b},c)}(\overline{x})$ and $K_1^{(q,a,\overline{b})}(\overline{x})$, where the asterisk as a superscript in the previous notation stands for the anti-wall of the involved solution. Notice that the configuration of four single domain walls that conform the $K_4^{(q,a,b,c)}(\overline{x},\gamma)$-solutions always involves the presence of a wall and its antiwall.

\FIGURE{\includegraphics[height=3cm]{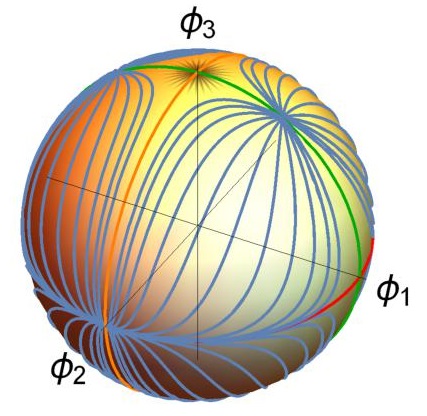}\qquad  
\includegraphics[height=3cm]{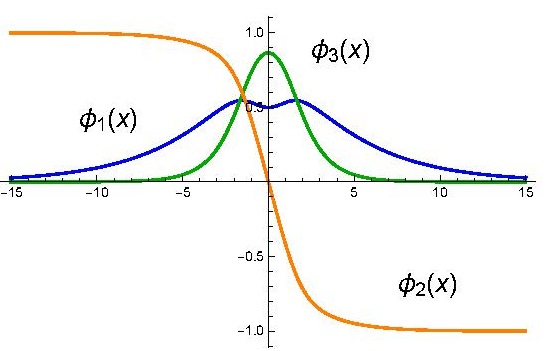} \qquad 
\includegraphics[height=3cm]{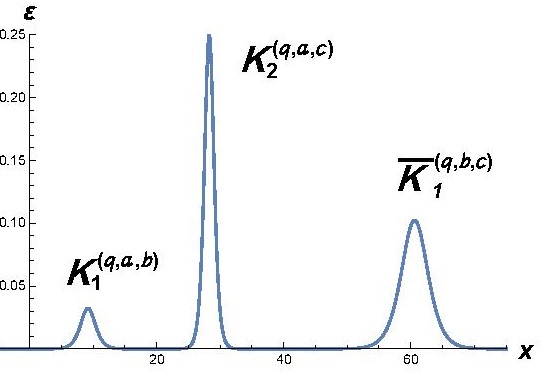}\caption{Orbits for several members of the family $K_4^{(q,a,b,c)}(\bar{x},\gamma)$ (left), field components (middle) and wall tension densities (right) graphically represented in Cartesian coordinates.}}

\noindent The following energy sum rules
\begin{eqnarray*}
\Omega [K_4^{(q,a,b,c)}(\bar{x},\gamma)] &=& \Omega[K_1^{(q,a,b)}(\overline{x})] + \Omega [K_2^{(q,a,c)}(x)] +\Omega[\overline{K}_1^{(q,b,c)}(\overline{x})] = \\ &=&  2 \, \Omega[K_1^{(q,a,b)}(\overline{x})] + 2 \, \Omega[\overline{K}_1^{(q,b,c)}(\overline{x})] =  R^3
\end{eqnarray*}
hold in this case. In summary, the family of $K_4^{(q,a,b,c)}(\bar{x},\gamma)$-domain walls can be understood as four solitonic thick 2-branes, two of them (of different type) equally separated from a superposition of the other two at a distance fixed by the value of the family parameter $\gamma$. It can be checked that the $\gamma \rightarrow \infty$ limits
\begin{eqnarray*}
&& \lim_{\tiny\begin{array}{c} \gamma\rightarrow \infty \\ x_0 \,\, {\rm constant} \end{array}}  K_4^{(q,a,b,c)}(x,\gamma) = \overline{K}_1^{(-q,b,c)}(x)  \hspace{0.3cm}, \\
&& \hspace{-0.4cm} \lim_{\tiny\begin{array}{c} \gamma \rightarrow \infty \\ \gamma+x_0 \equiv {\rm constant} \end{array}}  K_4^{(q,a,b,c)}(x,\gamma) = K_2^{(q,a,c)}(x) \hspace{0.3cm}, \\
&& \hspace{-0.5cm} \lim_{\tiny\begin{array}{c} \gamma\rightarrow \infty \\ \gamma+ \overline{\sigma}^2 x_0 \,\, {\rm constant} \end{array}}  K_4^{(q,a,b,c)}(x,\gamma) = K_1^{(q,a,(b+1){\rm mod}\,2)}(x)  \hspace{0.3cm},
\end{eqnarray*}
provide us with the singular domain walls. Similarly, the $\gamma \rightarrow -\infty$ limits lead to
\begin{eqnarray*}
&& \lim_{\tiny\begin{array}{c} \gamma\rightarrow -\infty \\ x_0 \,\, {\rm constant} \end{array}}  K_4^{(q,a,b,c)}(x,\gamma) = \overline{K}_1^{(q,(b+1){\rm mod}\,2,c)}(x)  \hspace{0.3cm}, \\
&& \hspace{-0.4cm} \lim_{\tiny\begin{array}{c} \gamma \rightarrow -\infty \\ \gamma+x_0 \equiv {\rm constant} \end{array}}  K_4^{(q,a,b,c)}(x,\gamma) = K_2^{(-q,a,c)}(x) \hspace{0.3cm}, \\
&& \hspace{-0.4cm} \lim_{\tiny\begin{array}{c} \gamma\rightarrow -\infty \\ \gamma+ \overline{\sigma}^2 x_0 \,\, {\rm constant} \end{array}}  K_4^{(q,a,b,c)}(x,\gamma) = K_1^{(-q,a,b)}(x)  \hspace{0.3cm}.
\end{eqnarray*}

\section{Linear stability of the domain walls}

In this section the linear stability of the domain walls described in the previous Section is investigated. The singular solutions were described by using spherical coordinates. In this context the domain wall solutions will follow the form $\Theta(x)=(\theta(x),\varphi(x))$. The study of the behavior of small fluctuations $\eta(x)=(\eta_1(x),\eta_2(x))$ around these domain walls requires the analysis of second-order differential operator \cite{Alonso2008b, Alonso2010}:
\begin{equation}
\Delta \eta\, =\, -\nabla_{\Theta'}\nabla_{\Theta'} \eta - R(\Theta',\eta) \Theta'-\nabla_\eta \, {\rm grad} \, U \hspace{0.3cm}, \label{hessian}
\end{equation}
which has been written in terms of covariant derivatives and the curvature tensor. Specifically, in the standard basis $\{ \frac{\partial}{\partial \theta}, \frac{\partial}{\partial \varphi} \}$ for the tangent space to the sphere along the domain wall, the vector fields are written as
\[
\Theta'(x) =\theta'(x) \frac{\partial}{\partial \theta} + \varphi'(x)\frac{\partial}{\partial \varphi}\ ;\quad \eta(x)=\eta_1  \frac{\partial}{\partial \theta} + \eta_2 \frac{\partial}{\partial \varphi}
\]
and $\nabla_\eta \,{\rm grad}\, U$ leads to the Hessian of the potential function. In general, the explicit form of this operator for a given domain wall is very complex. However, the special geometry of the singular domain wall orbits allows us to determine the complete spectrum for these singular solutions. The results are described in the following points:

\medskip

\noindent {\sl 1. Equatorial domain walls.} Plugging the expressions of Equatorial domain walls into (\ref{hessian}), we obtain the following small fluctuation operator:
\begin{eqnarray}
\Delta_{\rm Eq} \eta &=&  \left[ -\frac{d^2\eta_1}{dx^2}+\frac{R^2}{2} \left( 1+\sigma^4-\frac{3\bar{\sigma}^4}{2 \, \cosh^2(R\bar{\sigma}^2 x)}-(1-\sigma^4) \tanh(R\bar{\sigma}^2x)\right) \eta_1\right] \frac{\partial}{\partial \theta} \nonumber \\ && + \left[ -\frac{d^2\eta_2}{dx^2}+R^2 \bar{\sigma}^4  \left( 1-\frac{2}{\cosh^2(R\bar{\sigma}^2 x)}\right) \eta_2\right] \frac{\partial}{\partial \varphi} \hspace{0.3cm}. \label{hess1}
\end{eqnarray}
The first component of (\ref{hess1})
\[
{\cal H}_{11} = - \frac{d^2}{dx^2} + \frac{R^2}{2} \Big[ 1+\sigma^4 - \frac{3 \bar{\sigma}^4 }{2 \cosh^2(R \bar{\sigma}^2 x)} - (1-\sigma^4) \tanh(R\bar{\sigma}^2 x) \Big]
\]
governs the behavior of the orthogonal fluctuations around the $K_1^{(q,a,b)}(\overline{x})$-domain walls. The spectrum of ${\cal H}_{11}$ consists of a continuous spectrum which is simply degenerate in the range $[\sigma^4 R^2,R^2]$ and doubly degenerate for eigenvalues in the interval $(R^2,\infty)$. On the other hand, the small longitudinal fluctuation operator
\[
{\cal H}_{22} = - \frac{d^2}{dx^2} + R^2 \bar{\sigma}^4 \Big( 1- \frac{2}{\cosh^2(R \bar{\sigma}^2 x)} \Big)
\]
involves the presence of a zero mode $\omega_0^2=0$ and a doubly degenerate continuous spectrum emerging on the threshold value $\omega^2=R^2 \bar{\sigma}^4$. The existence of a longitudinal zero mode is a general rule and establishes that the domain wall center can be set at any spatial point $x_0\in \mathbb{R}$. The lack of negative eigenvalues implies that the $K_1^{(q,a,b)}(\overline{x})$-domain walls are stable.

\medskip

\noindent {\sl 2. Meridian domain walls.} The $\left(\pm \frac{\pi}{2}\right)$-Meridian domain wall fluctuation operator is given by the expression
\begin{eqnarray*}
&& \hspace{-0.5cm}\Delta_{\rm Mer} \eta=  \left( -\frac{d^2\eta_1}{dx^2}+R^2 \sigma^4  \left( 1-\frac{2}{\cosh^2(R \sigma^2 x)}\right) \eta_1\right) \frac{\partial}{\partial \theta} \\ && + \left( -\frac{d^2\eta_2}{dx^2}-R\sigma^2\left( 1-\tanh(R\sigma^2x)\right) \frac{d\eta_2}{dx} + R^2 \bar{\sigma}^2  \left( 1-\sigma^2\tanh (R\sigma^2x) \right) \eta_2\right) \frac{\partial}{\partial \varphi} \hspace{0.4cm} .
\end{eqnarray*}
The form of this operator can be simplified if a basis for the tangent space that is transported in a parallel way along the domain wall is considered \cite{Alonso2008b, Alonso2010}. The parallel transport equations along Meridian solutions for a generic vector field $v(x) = v^1(x) \frac{\partial }{\partial \theta} + v^2(x)\frac{\partial }{\partial \theta}$ read as
$\frac{dv^1}{dx} =0$ and $\frac{dv^2}{dx}= -\frac{R \sigma^2 v^2}{e^{2 R \sigma^2 x}+1}$, which leads to the solution $v^1(x) =1$ and $v^2(x)=\sqrt{1+e^{-2R\sigma^2 x}}$. Therefore,
\[
\left\{ v_1= \frac{\partial}{\partial \theta}, v_2= \sqrt{1+e^{-2R\sigma^2 x}}\, \frac{\partial}{\partial \varphi}\right\}
\]
is a parallel frame along $\left(\pm \frac{\pi}{2}\right)$-Meridian solutions. Writing the deformation field in this basis: $\bar{\eta} =\bar{\eta}_1 \, v_1+\bar{\eta}_2 \, v_2$, the Hessian operator reads:
\begin{eqnarray*}
&& \hspace{-0.5cm} \Delta_{\rm Mer} \bar{\eta}=  \left[ -\frac{d^2\eta_1}{dx^2}+R^2 \sigma^4  \left( 1-\frac{2}{\cosh^2(R \sigma^2 x)}\right) \eta_1\right]\, v_1 +\\ && + \left[ -\frac{d^2\eta_2}{dx^2}+\frac{R^2}{2} \left( \sigma^4-2\sigma^2+2 -\frac{3\sigma^4}{2\cosh^2(R\sigma^2 x)}- \sigma^2(2-\sigma^2) \tanh(R\sigma^2x)\right)  \bar{\eta}_2\right] v_2
\end{eqnarray*}
For these solutions the longitudinal fluctuation operator
\[
{\cal H}_{11} = - \frac{d^2}{dx^2} + R^2 \sigma^4 \Big( 1  - \frac{2}{\cosh^2(R \sigma^2 x)} \Big)
\]
comprises a zero mode $\omega_0^2=0$ together with a doubly degenerate spectrum $\omega^2 \in (R^2\sigma^4, \infty)$. The spectrum of the second component
\[
{\cal H}_{22} = - \frac{d^2}{dx^2} + \frac{R^2}{2} \Big(\sigma^4-2 \sigma^2+2 - \frac{3\sigma^4}{2\cosh^2(R \sigma^2 x)} -\sigma^2(2-\sigma^2) \tanh(R \sigma^2 x) \Big)
\]
is simply degenerate in the interval $\omega^2\in [\bar{\sigma}^4  R^2,R^2]$ and doubly degenerate in $\omega^2\in (R^2,\infty)$. From the previous analysis, we conclude that the $\overline{K}_1^{(q,b,c)}(\overline{x})$-solutions are stable topological domain walls.

\medskip

\noindent {\sl 3. Prime Meridian domain walls.} This case is similar to the previous one, the $K_2^{(q,a,c)}(\overline{x})$-fluctuation operator extracted from (\ref{hessian}) is given by:
\begin{eqnarray*}
&& \hspace{-0.5cm} \Delta_{\rm PMer} \eta=  \left( -\frac{d^2\eta_1}{dx^2}+R^2   \left( 1-\frac{2}{\cosh^2(R  x)}\right) \eta_1\right) \frac{\partial}{\partial \theta} \\ && + \left( -\frac{d^2\eta_2}{dx^2}-R\left( 1-\tanh(R x)\right) \frac{d\eta_2}{dx} - R^2 \bar{\sigma}^2  \left( \sigma^2 - \tanh (R x) \right) \eta_2\right) \frac{\partial}{\partial \varphi}
\end{eqnarray*}
in the standard frame. Now the parallel frame is defined by the system
\[
\left\{ v_1= \frac{\partial}{\partial \theta}, v_2= \sqrt{1+e^{-2R x}}\, \frac{\partial}{\partial \varphi}\right\}
\]
which is obtained by solving the parallel transport equations $\frac{dv^1}{dx} =0$ and $\frac{dv^2}{dx}=  -\frac{R v^2}{e^{2 R x}+1}$. In this frame the previous operator reads
\begin{eqnarray*}
&& \hspace{-0.5cm} \Delta_{\rm PMer} \bar{\eta}=  \left[ -\frac{d^2\eta_1}{dx^2}+R^2  \left( 1-\frac{2}{\cosh^2(R  x)}\right) \eta_1\right] \, v_1 +\\ && +  \left[ -\frac{d^2\eta_2}{dx^2}+\frac{R^2}{2} \left( 2\sigma^4-2\sigma^2+1 -\frac{3}{2\cosh^2(R x)}+ (1-2\sigma^2) \tanh(R x)\right)  \bar{\eta}_2 \right] v_2
\end{eqnarray*}
The spectrum of this operator comprises two zero modes, each of them associated with the two different components of $\Delta_{\rm PMer}$. A doubly degenerate continuous spectrum emerges at the threshold value $\omega^2=R^2$ for the first component. For the second component a simply and doubly degenerate spectra in the intervals $[ \min\{\bar{\sigma}^4  R^2,\sigma^4 R^2\},\max \{\bar{\sigma}^4  R^2,\sigma^4 R^2\}  ]$ and $(\max \{\bar{\sigma}^4  R^2,\sigma^4 R^2\},\infty)$ arise. As a consequence the $K_2^{(q,a,c)}(\overline{x})$-domain walls are stable although a new neutral stability channel is open for these solutions.

Finally, Morse Theory can be applied to the domain wall orbit space to establish the stability of the domain wall families $K_2^{(q,a,b,c)}(\overline{x},\gamma)$ and $K_4^{(q,a,b,c)}(\overline{x},\gamma)$. In the first case the set of $K_2^{(q,a,b,c)}(\overline{x},\gamma)$-orbits lacks conjugate points, which implies that these solutions are stable. We recall that every member of this family can be interpreted as a non-linear combination of the single walls $K_1^{(q,a,b)}(\overline{x})$ and $\overline{K}_1^{(q,b,c)}(\overline{x})$, which form a stable configuration. On the other hand, all the $K_4^{(q,a,b,c)}(\overline{x},\gamma)$ orbits cross through the conjugate point $F_{ac}$. This allows us to conclude that these solutions are unstable. Recall that in the description of this family was noted that these solutions involve the presence of a wall and its anti-wall, which obviously is a source of unstability, confirmed in this case by the application of Morse theory.

\section{Conclusions and further comments}

In this paper we have analytically calculated all the static domain wall variety which arises in a non-linear $\mathbb{S}^2$-sigma model with a homogeneous quartic polynomial potential. The vacuum set in this model involves the existence of six maximally separated points on the sphere. The pieces of great circles joining these points split the sphere in eight octants and are the orbits of the so called Equatorial, $(\pm \frac{\pi}{2})$-Meridian and Primer Meridian domain walls. The first two types describe two basic domain walls, whose tension density is localized around one point. The Prime Meridian domain walls appear when a Equatorial domain wall and a $(\pm \frac{\pi}{2})$-Meridian domain wall are exactly overlapped. Indeed, these solutions are singular members of the $K_2^{(q,a,b,c)}(\overline{x},\gamma)$-domain wall families. The members of these families are characterized as a non-linear combination of the two previously mentioned basic domain walls, which are separated by a distance parameterized by the value of $\gamma$. The energy density of these solutions are condensed around two points. No interactions between these walls arise when they remain still. All these solutions have been proved to be stable domain walls. A similar equilibrium configuration is obtained when four basic domain walls (two of each type) are alternately disposed along the space if the two walls in the middle are exactly overlapped and equally separated of the other two. These static solutions constitute the $K_4^{(q,a,b,c)}(\overline{x},\gamma)$-domain wall families described in the paper. These complex internal structure makes these topological defects into unstable.

As previously noted, there is no forces acting between the basic domain walls when they are motionless. Indeed, all the members in the $K_2^{(q,a,b,c)}(\overline{x},\gamma)$-families are energy degenerate. Obviously this picture takes place at classical level. From our point of view, it is very interesting to investigate whether this situation persists in the quantum realm. One-loop effects in field theory can be quantified by using the Gilkey-De Witt-Avramidi heat kernel expansion approach \cite{Vassilevich2003, Avramidi2002, Alonso2002e}. This technique must be modified when zero mode fluctuations are involved, as in our case. This problem has been successfully tackled in the scalar field theory \cite{Alonso2012, Alonso2013} and even in the Abelian Higgs model \cite{Alonso2016b,Alonso2016}. As a final remark we conjecture the presence of quantum-induced interactions in the moduli space of the previous degenerate BPS domain walls in concordance with other models, see \cite{Alonso2014, Alonso2004, Alonso2002d}.

\section{ACKNOWLEDGEMENTS}

The authors acknowledge the Spanish Ministerio de Econom\'{\i}a y Competitividad for financial support under grant MTM2014-57129-C2-1-P. They are also grateful to the Junta de Castilla y Le\'on for financial help under grant VA057U16.

\end{document}